\documentclass[journal]{IEEEtran}
\usepackage{amsmath,amssymb,amsfonts}
\usepackage{dcolumn}
\usepackage{algorithmic}
\usepackage{graphicx}
\usepackage{booktabs}
\usepackage{multirow}
\usepackage{xspace}
\usepackage{url}
\usepackage{color}
\graphicspath{ {./figures/} }
\newcommand{\mat}[1]{\boldsymbol{#1}} 
\newcommand{\methodname}{{DU-GAN}\xspace}

\newcommand{\ie}{\textit{i}.\textit{e}.\xspace}

\usepackage{hyperref}

\makeatother

\begin{document}

\title{\methodname: Generative Adversarial Networks with Dual-Domain U-Net Based Discriminators for Low-Dose CT Denoising}

\author{
\thanks{© 2021 IEEE. Personal use of this material is permitted. Permission from IEEE must be
obtained for all other uses, in any current or future media, including
reprinting/republishing this material for advertising or promotional purposes, creating new
collective works, for resale or redistribution to servers or lists, or reuse of any copyrighted
component of this work in other works. (\textit{Corresponding author: Hongming Shan})}
\thanks{
Z. Huang and J. Zhang are with the Shanghai Key Lab of Intelligent Information Processing and the School of Computer Science, Fudan University, Shanghai 200433, China (email: zzhuang19@fudan.edu.cn;  jpzhang@fudan.edu.cn).}
\thanks{
Y. Zhang is with the College of Computer Science, Sichuan University, Chengdu
610065, China  (e-mail: yzhang@scu.edu.cn).
}
\thanks{
H. Shan is with the Institute of Science and Technology for Brain-inspired Intelligence and  MOE Frontiers Center for Brain Science, Fudan University, Shanghai, 200433, China, and also with the Shanghai Center for Brain Science and Brain-inspired Technology, Shanghai 201210, China (e-mail: hmshan@fudan.edu.cn).
}
Zhizhong Huang,~\IEEEmembership{Graduate Student Member,~IEEE},
Junping Zhang,~\IEEEmembership{Member,~IEEE},\\
Yi Zhang,~\IEEEmembership{Senior Member,~IEEE}, and
Hongming Shan,~\IEEEmembership{Member,~IEEE}
}

\markboth{IEEE Transactions on Instrumentation and Measurement, vol. xx, 2021}%
{Shell \MakeLowercase{\textit{et al.}}: Bare Demo of IEEEtran.cls for IEEE Journals}

\maketitle

\begin{abstract}
Low-dose computed tomography~(LDCT) has drawn major attention in the medical imaging field due to the potential health risks of CT-associated X-ray radiation to patients. Reducing the radiation dose, however, decreases the quality of the reconstructed images, which consequently compromises the diagnostic performance. Over the past few years, various deep learning techniques, especially generative adversarial networks (GANs), have been introduced to improve the image quality of LDCT images through denoising, achieving impressive results over traditional approaches. GANs-based denoising methods usually leverage an additional classification network, \ie discriminator, to learn the most discriminate difference between the denoised and normal-dose images and, hence, regularize the denoising model accordingly; it often focuses either on the global structure or local details. To better regularize the LDCT denoising model, this paper proposes a novel method, termed \methodname, which leverages U-Net based discriminators in the GANs framework to learn both global and local difference between the denoised and normal-dose images in both image and gradient domains. The merit of such a U-Net based discriminator is that it can not only provide the per-pixel feedback to the denoising network through the outputs of the U-Net but also focus on the global structure in a semantic level through the middle layer of the U-Net. In addition to the adversarial training in the image domain, we also apply another U-Net based discriminator in the image gradient domain to alleviate the artifacts caused by photon starvation and enhance the edge of the denoised CT images. Furthermore, the CutMix technique enables the per-pixel outputs of the U-Net based discriminator to provide radiologists with a confidence map to visualize the uncertainty of the denoised results, facilitating the LDCT-based screening and diagnosis. Extensive experiments on the simulated and real-world datasets demonstrate superior performance over recently published methods both qualitatively and quantitatively. Our source code is made available at \url{https://github.com/Hzzone/DU-GAN}.

\end{abstract}


\section{Introduction}

Computed tomography~(CT) can provide the cross-sectional images of the internal body by the x-ray radiation, which is one of the most important imaging modalities in clinical diagnosis. Although CT plays an essential role in diagnosing diseases, the widespread use of CT is raising more and more public concerns towards its safety since CT-related X-ray radiation may cause unavoidable damage to the health of humans and induce cancers. Consequently, reducing the radiation dose of CT as low as reasonably achievable (\emph{a.k.a.} ALARA) is a well-accepted principle in CT-related research over the past decades~\cite{shah2008alara}. The reduction of radiation dose, however, inevitably brings the noise and artifacts into the reconstructed images, severely compromising the subsequent diagnosis and other tasks such as LDCT-based lung nodule classification~\cite{lei2020shape}. 

A straightforward way to address this issue is to reduce the noise in the LDCT image~\cite{attivissimo2010technique,wang2020deep}. However, it remains a challenging problem due to its ill-posed nature. 
In recent years, various deep learning based methods have been proposed for LDCT denoising~\cite{chen2017cnn,chen2017low,shan20183,yang2018low,shan2019competitive,wolterink2017generative,wang2018image}, achieving impressive results. There are two key components in designing a denoising model: network architecture and loss function; the former one can determine the capacity of the denoising model while the latter one can control how the denoised images visually look like~\cite{shan20183}. Although different network architectures  such as 2D convolutional neural networks (CNNs)~\cite{chen2017cnn}, 3D CNNs~\cite{shan20183,wolterink2017generative}, and residual encoder-decoder CNNs~(RED-CNN)~\cite{wu2017cascaded} have been explored for LDCT denoising, literature has shown that the loss function is relatively more important than the network architecture as it has a direct impact on the image quality~\cite{shan20183,zhao2016loss}. 

One of the most popular loss functions is the mean-squared error (MSE), which computes the average of the squares of the per-pixel errors between the denoised and normal-dose images. Although gaining impressive performance in terms of peak signal-to-noise (PSNR), MSE usually leads to over-smoothened images, which has been proven to poorly correlate with the human perception of image quality~\cite{goodfellow2014generative,wang2004image}. In view of this observation, alternative loss functions such as perceptual loss, $\ell_1$ loss, and adversarial loss have been investigated for LDCT denoising. Among them, adversarial loss has been shown to be a powerful one as it can dynamically measure the similarity between the denoised and normal-dose images during the training, which enables the denoised images to preserve more texture information from normal-dose one. The computation of the adversarial loss is based on the discriminator, which is a classification network to learn a representation differentiating the denoised images from the normal-dose images; it can measure the most discriminant difference either in a global or local level, depending on that one unit of the output of discriminator corresponds to the whole image or a local region. Such a discriminator is prone to forgetting previous difference because the distribution of synthetics samples shifts as the generator constantly changes through training, failing to maintain a powerful data representation to characterize the global and local image difference~\cite{schonfeld2020u}. As a result, it often results in the generated images with discontinued and mottled local structures~\cite{lin2019coco} or images with incoherent geometric and structural patterns~\cite{zhang2019self}. In addition to the noise, LDCT images may contain severe streak artifacts caused by photon starvation, which may not be effectively removed through the loss function solely in the image domain.

To learn a powerful data representation to regularize the denoising model in the adversarial training, we propose a U-Net~\cite{ronneberger2015u} based discriminator in the GANs framework for LDCT denoising, termed \methodname, which can simultaneously learn the global and local difference between the denoised and normal-dose images in image and gradient domains. More specifically, our proposed discriminator follows the U-Net architecture including an encoder and a decoder network, where the encoder encodes the input to a scalar value focusing on the global structures while the decoder reconstructs a per-pixel confidence map capturing the changes of local details between the denoised and normal-dose images. In doing so, it can provide not only the per-pixel feedback but also the global structural difference to the denoising network. In addition to the adversarial training in the image domain, we also apply another U-Net based discriminator in the image gradient domain to alleviate the artifacts caused by photon starvation and enhance the edge of the denoised images. Moreover, to regularize the U-Net based discriminator, we introduce the CutMix data augmentation to mix the denoised and normal-dose images. Consequently, the U-Net based discriminator can provide radiologists with the per-pixel outputs as a confidence map to visualize the uncertainty of the denoised results, which can facilitate radiologists' screening and diagnosis when using the denoised LDCT images.

The benefits of the proposed \methodname are as follows.
\begin{enumerate}
    \item Unlike existing GAN-based denoising methods that use a classification as the discriminator, the proposed \methodname utilizes a U-Net based discriminator for LDCT denoising, which can simultaneously learn global and local difference between the denoised and normal-dose images. Consequently, it can provide not only the per-pixel feedback but also the global structural difference to the denoising model.

    \item In addition to adversarial training in the image domain, the proposed \methodname also performs adversarial training in the image gradient domains, which can alleviate the streak artifacts caused by photon starvation and enhance the edge of the denoised images.

    \item The proposed \methodname can provide radiologists with a confidence map visualizing the uncertainty of the denoised results through the CutMix technique, which could facilitate radiologists' screening and diagnosis when using the denoised LDCT images.
    
    \item Extensive experiments on simulated and real-world datasets demonstrate the effectiveness of the proposed method through both qualitative and quantitative comparisons.
\end{enumerate}

The remainder of this paper is organized as follows. We briefly survey the developments of the LDCT denoising methods and generative adversarial networks in Section~\ref{related_work}. We present our LDCT denoising framework \methodname with dual-domain U-Net based discriminators, and then introduce the CutMix regularization technique as well as the network architectures and loss functions in our framework in Section~\ref{sec:method}, followed by both qualitative and quantitative comparisons with the state-of-the-art methods on the simulated and real-world datasets in Section~\ref{sec:exp}. Finally, we conclude this paper in Section~\ref{sec:conc}.

\section{Related Work}
\label{related_work}

This section briefly surveys the development of LDCT denoising and generative adversarial networks.

\subsection{LDCT Denoising}

The noise reduction algorithms for LDCT can be summarized into three categories: 1) sinogram filtration; 2) iterative reconstruction; and 3) image post-processing. As a significant difference from routine CT, the LDCT acquires noisy sinogram data from scanner. A straightforward solution is to perform the denoising process on the sinogram data before image reconstruction, \ie sinogram filtration-based methods~\cite{wang2005sinogram,wang2006penalized,manduca2009projection}. Iterative reconstruction methods combine the statistics of raw data in the sinogram domain~\cite{ramani2011splitting,wu2021tensor} and the prior information in the image domain such as total variation~\cite{zheng2018pwls} and dictionary learning~\cite{xu2012low}; these pieces of generic information can be effectively integrated in the maximum likelihood and compressed sensing frameworks. These two categories, however, require the access to raw data that are typically unavailable from commercial CT scanner. 

\begin{figure*}[t]
    \centering
    \includegraphics[width=1.0\linewidth]{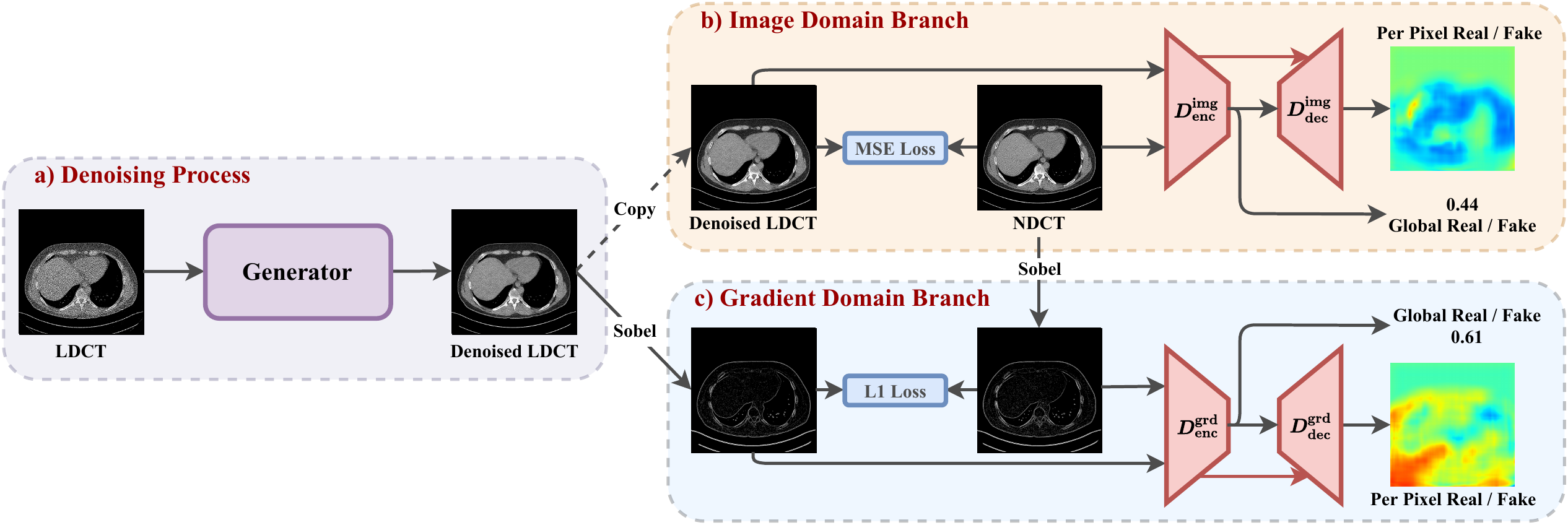}
    \caption{Overall framework of our proposed \methodname. The generator produces denoised LDCT images, and two independent branches with U-Net based discriminators perform at the image and gradient domains. The U-Net based discriminator provides both global structure and local per-pixel feedback to the generator. Furthermore, the image discriminator encourages the generator to produce photo-realistic CT images while the gradient discriminator is utilized for better edge and alleviating streak artifacts caused by photon starvation.}
    \label{fig:framework}
\end{figure*}

Different from the previous two categories, image post-processing methods directly operate on the reconstructed images that are publicly available after removing patient privacy. Traditional methods such as non-local means~\cite{ma2011low} and block-matching 3D~\cite{feruglio2010block}, however, lead to the loss of some critical structural details and result in over-smoothened denoised LDCT images. 
The rapid development of deep learning techniques has advanced many medical applications.
In LDCT denoising, deep-learning-based models have achieved impressive results~\cite{chen2017cnn,shan20183,shan2019competitive,wolterink2017generative,wu2017cascaded,li2020sacnn}. There are two critical components in designing a deep-learning-based denoising model: network architecture and loss function; the former one determines the capacity of a denoising model while the later one controls how the denoised images visually look like. Although the literature has proposed several different network architectures for LDCT denoising such as 2D CNNs~\cite{chen2017cnn}, 3D CNN~\cite{shan20183,wolterink2017generative}, RED-CNN~\cite{chen2017cnn}, and cascaded CNN~\cite{wu2017cascaded}, the literature has shown that the loss function plays a relatively more important role than network architecture as it has a direct impact on the image quality~\cite{shan20183,zhao2016loss}. The simplest loss function is the MSE, which however has been shown to poorly correlate with the human perception of image quality~\cite{goodfellow2014generative,wang2004image}.  In view of this observation, alternative loss functions such as perceptual loss, $\ell_1$ loss, adversarial loss, or mixed loss functions have been investigated for LDCT denoising.  Among them, adversarial loss has been shown to be a powerful one as it can dynamically measure the similarity between the denoised and normal-dose images during the training, which enables the denoised images to preserve more texture information from normal-dose one. Adversarial loss reflects either global or local similarity, depending on the design of discriminator.

Unlike the conventional adversarial loss, the adversarial loss used in this study is based on a U-Net based discriminator, which can simultaneously characterize global and local difference between the denoised and normal-dose images, better regularizing the denoising model. That is, DU-GAN enjoys both advantages of the per-pixel discriminator capturing the changes at pixel level and traditional classification discriminator focusing on global structures. In addition to the adversarial loss in the image domain, the adversarial loss in the image gradient domain proposed in this paper can alleviate the streak artifacts caused by photon starvation and enhance the edge of the denoised images.

\subsection{Generative Adversarial Networks (GANs)}
As one of the most hot research topics in recent years, GANs~\cite{goodfellow2014generative} and their variants have been successfully applied to various tasks~\cite{karras2017progressive,isola2017image,guo2020unsupervised}. They typically consist of two networks: 1) a generator learning to capture the data distribution of training data and produce new samples that are indistinguishable from the real ones, and 2) a discriminator attempting to distinguish real samples from fake ones produced by the generator.
These two networks are trained alternatively, ending once the balance is achieved. In the context of LDCT denoising, the generator aims to produce photo-realistic denoised results to fool the discriminator while the discriminator tries to distinguish the real normal-dose CT~(NDCT) images and denoised ones.  To foster the stability of training GANs, various variants of GANs have been proposed, such as Wasserstein GAN~(WGAN)~\cite{arjovsky2017wasserstein}, WGAN with gradient penalty (WGAN-GP)~\cite{gulrajani2017improved}, and least-squares GANs~\cite{mao2017least}.

In this paper, we adopt the least-squares GANs~\cite{mao2017least}, spectral normalization~\cite{miyato2018spectral}, and U-Net based discriminator~\cite{schonfeld2020u} to form the GANs framework for LDCT denoising. As a significant difference, our \methodname  performs adversarial training in both image and gradient domains, which can reduce noise and alleviate streak artifacts simultaneously. We note that the proposed \methodname is also suitable for other variants of GAN such as WGAN and WGAN-GP.

\section{Methodology}\label{sec:method}

Fig.~\ref{fig:framework} presents the proposed \methodname for LDCT denoising, which contains a denoising model as generator, and two U-Net based discriminators in both image and gradient domains. We highlight that the U-Net based discriminator is able to learn the global and local difference between denoised and normal-dose images. Next, we present all components, network architecture, and loss functions in detail, followed by its complexity.

\subsection{The Denoising Process}
\label{sec:model}

The denoising process is to learn a generative model $G$ that maps an LDCT image $\mat{I}_{\mathrm{LD}} \in \mathbb{R}^{w \times h}$ of size $w \times h$ to its normal-dose CT~(NDCT) counterpart $\mat{I}_{\mathrm{ND}} \in \mathbb{R}^{w \times h}$ by removing the noise in LDCT image. Formally, it can be written as:
\begin{align}
    \mat{I}_{\mathrm{den}} = G(\mat{I}_{\mathrm{LD}})\approx\mat{I}_{\mathrm{ND}}, 
\end{align}
where $\mat{I}_{\mathrm{den}}$ denotes the denoised LDCT image. Typically, LDCT denoising can be seen as a specific image translation problem. Therefore, the GANs-based methods~\cite{shan20183,yang2018low,shan2019competitive,yi2018sharpness} utilize the GANs to improve the visual quality of denoised LDCT images thanks to its strong capability of GANs in generating high-quality images. Different from the conventional GANs that take a noise vector to generate an image, our denoising model serves as the generator that only takes the LDCT image as the input. In this study, we used the RED-CNN~\cite{chen2017low} as the denoising model to demonstrate the effectiveness of the dual-domain U-Net based discriminators in the adversarial training.

\begin{figure}[t]
    \centering
    \includegraphics[width=1\linewidth]{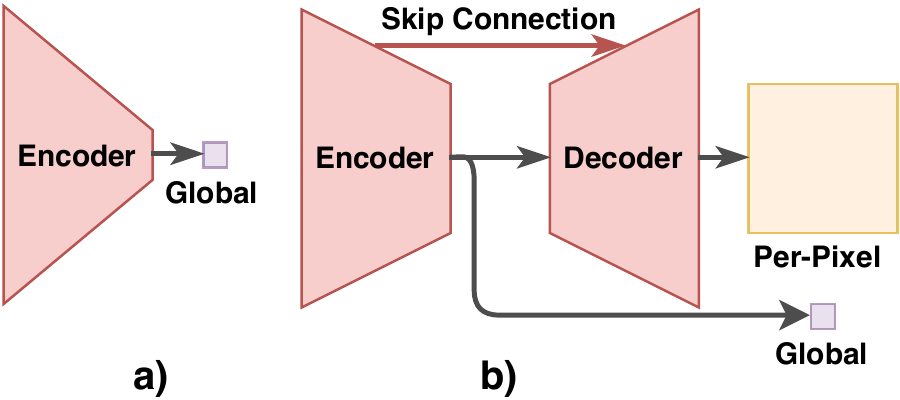}
    \caption{The difference between a) traditional classification discriminator, and b) U-Net based discriminator.  U-Net based discriminator extends traditional one to capture the global and local information simultaneously.}
    \label{fig:discriminators}
\end{figure}
\subsection{Dual-Domain U-Net Based Discriminator}
\label{sec:unetD}

The GANs-based methods~\cite{shan20183,yang2018low,shan2019competitive,yi2018sharpness} for LDCT denoising usually maintain the competition of GANs under the structural level, whose discriminator progressively downsamples the input into a scalar value and are trained with Wasserstein GANs~\cite{arjovsky2017wasserstein,gulrajani2017improved}, as shown in Fig.~\ref{fig:discriminators}(a). However, the discriminator is prone to forgetting previous samples because the distribution of synthetics samples shifts as the generator constantly changes during training, failing to maintain a powerful data representation to characterize the global and local image difference~\cite{schonfeld2020u,chen2019self}. 

To address the problems above, we introduce the U-Net based discriminators in both image and gradient domains.

\subsubsection{U-Net based discriminator in the image domain} To learn a powerful data representation that can characterize both global and local difference, we design an LDCT denoising framework based on GANs to deal with LDCT denoising. Traditionally, U-Net contains an encoder, a decoder, and several skip connections copying the feature-maps from the encoder to the decoder to preserve high-resolution features, which has demonstrated its state-of-the-art performance in many semantic segmentation tasks~\cite{zhou2018unet,han2018framing} and image translation tasks~\cite{schonfeld2020u,isola2017image}.
In the context of LDCT denoising, we highlight that U-Net and its variants are only used as the denoising model, which have not been explored as the discriminator. We adopt the U-Net to replace the standard classification discriminator in GANs to have a U-Net style discriminator that allows the discriminator to maintain both global and local data representation. Fig.~\ref{fig:discriminators}(b) details the architecture of U-Net based discriminator.

\begin{figure}[t]
    \centering
    \includegraphics[width=1\linewidth]{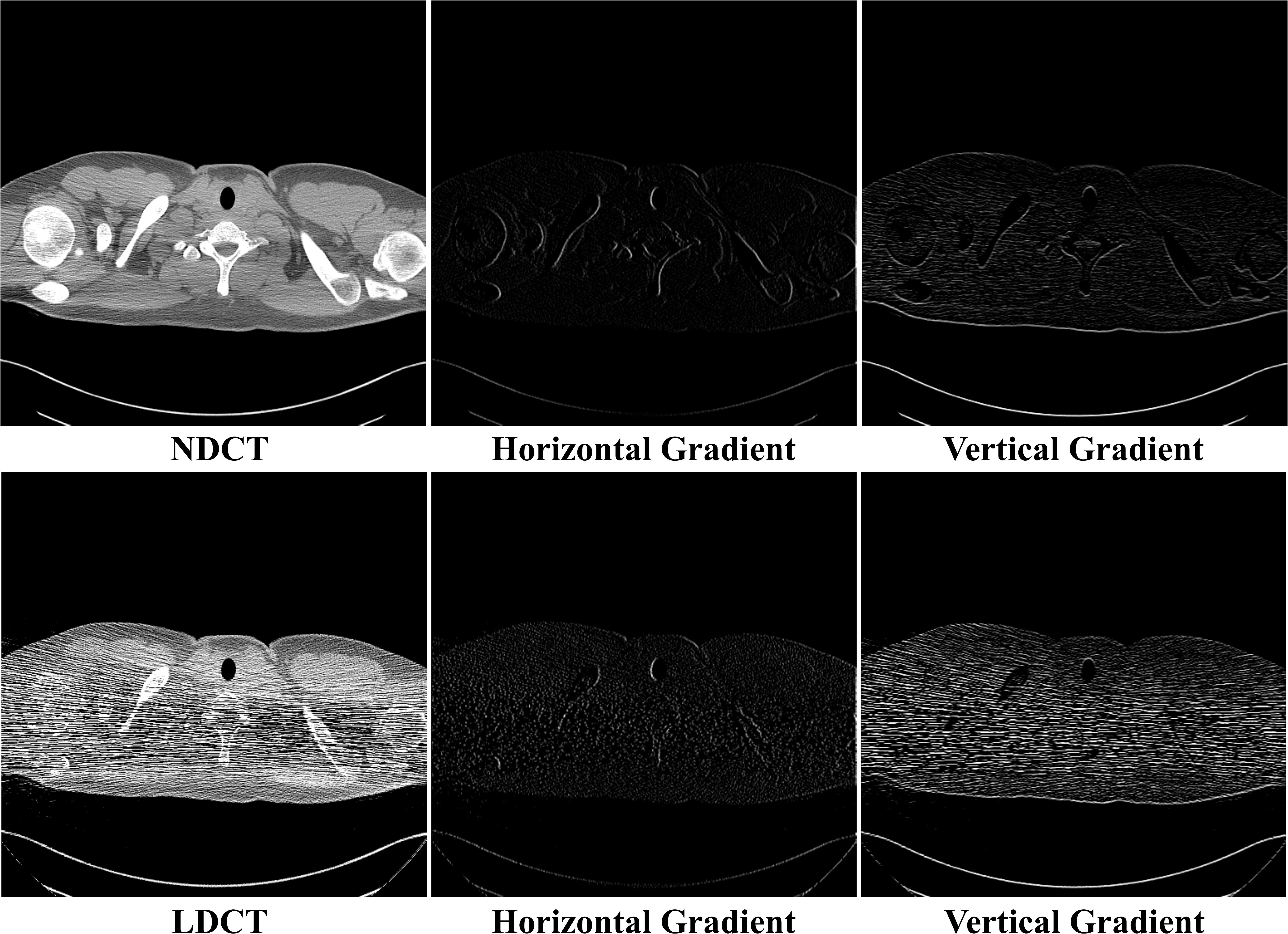}
    \caption{Gradients of horizontal and vertical directions from a pair of LDCT and NDCT images. The streak artifacts can be easily captured in the gradient domain.}
    \label{fig:gradient_example}
\end{figure}

Here, we use $D^{\mathrm{img}}$ to denote the U-Net based discriminator in the image domain. The encoder of $D^{\mathrm{img}}$, $D^{\mathrm{img}}_{\mathrm{enc}}$, follows the traditional discriminator that progressively downsamples the input using several convolutional layers, capturing the global structure context. On the other hand, the decoder $D^{\mathrm{img}}_{\mathrm{dec}}$ performs progressive upsampling with skip connections from encoder $D^{\mathrm{img}}_{\mathrm{enc}}$ in a reverse order, further enhancing the ability of discriminator to draw the local details of real and fake samples. Furthermore, the discriminator loss is computed from the outputs of both $D^{\mathrm{img}}_{\mathrm{enc}}$ and $D^{\mathrm{img}}_{\mathrm{dec}}$, while the traditional discriminator used in previous works~\cite{shan20183,yang2018low,yi2018sharpness} \emph{only} classifies the inputs into being real and fake from the encoder. In doing so, the U-Net based discriminator can provide more informative feedback to the generator including both local per-pixel and global structural information. In this paper, we employ the least-squares GANs~\cite{mao2017least} rather than conventional GANs~\cite{goodfellow2014generative} for the discriminators to stabilize the training process and improve the visual quality of denoised LDCT. Formally, the discriminator loss for $D^{\mathrm{img}}$ from both $D^{\mathrm{img}}_{\mathrm{enc}}$ and $D^{\mathrm{img}}_{\mathrm{dec}}$ can be written as:
\begin{align}
    \mathcal{L}_{D^{\mathrm{img}}} =& \underbrace{\mathbb{E}_{\mat{I}_{\mathrm{ND}}} \Big[D^{\mathrm{img}}_{\mathrm{enc}}\big(\mat{I}_{\mathrm{ND}}\big)-1\Big]^{2} + \mathbb{E}_{\mat{I}_{\mathrm{LD}}} \Big[D^{\mathrm{img}}_{\mathrm{enc}}\big(\mat{I}_{\mathrm{den}}\big)\Big]^{2}}_{\text{global\ adversarial}}+\notag\\
    &\underbrace{\mathbb{E}_{\mat{I}_{\mathrm{ND}}} \Big[D^{\mathrm{img}}_{\mathrm{dec}}\big(\mat{I}_{\mathrm{ND}}\big)-1\Big]^{2} + \mathbb{E}_{\mat{I}_{\mathrm{LD}}} \Big[D^{\mathrm{img}}_{\mathrm{dec}}\big(\mat{I}_{\mathrm{den}}\big)\Big]^{2}}_{\text{local\ adversarial}}\label{eq:dis_img}
\end{align}
where 1 is the decision boundary of least-squares GANs. 

\subsubsection{U-Net based discriminator in the gradient domain}
However, the competition in the image domain alone is only able to force the generator towards generating photo-realistic denoised LDCT images; it is insufficient to encourage better edge for keeping the pathological changes of original NDCT images and alleviate the streak artifacts caused by photon starvation in LDCT. Previous methods such as~\cite{shan2019competitive} measure the different MSE in the gradient domain, which may be insufficient to enhance the edge as MSE tends to blur image. 
To this end, we propose to perform an additional GANs competition in the gradient domain, where our motivation is presented in Fig.~\ref{fig:gradient_example}. Specifically, the streaks and edge in CT images are highlighted in their horizontal and vertical gradient magnitudes. Therefore, another branch of the gradients estimated by a Sobel operator~\cite{kanopoulos1988design} is performed aside the image branch, which encourages better edge information and alleviates streak artifacts. Similar to~\eqref{eq:dis_img}, we can define the discriminator loss  in the gradient domain $\mathcal{L}_{D^{\mathrm{grd}}}$, where $D^{\mathrm{grd}}$ represents the discriminator in the gradient domain. 

\subsubsection{Dual-domain U-Net based discriminators} 
Combining the U-Net based discriminators in the image and gradient domains, two independent GANs competitions are maintained during training. The overall framework of our proposed LDCT denoising model is shown in Fig.~\ref{fig:framework}. In detail, the generator is to denoise an LDCT image, which is then fed into two independent discriminators operating in the image and gradient domains. The discriminator $D^{\mathrm{img}}$ in the image domain branch penalizes the generator generating photo-realistic denoised LDCT while the discriminator $D^{\mathrm{grd}}$ in the gradient domain branch encourages better edge while alleviating streak artifacts caused by photon starvation. Additionally, the discriminator in each branch employs a U-Net based architecture to encourage the generator focusing both global structure and local details, which can also boost the interpretability of the denoising process with the per-pixel confidence map output by $D^{\mathrm{img}}_{\mathrm{dec}}$ and $D^{\mathrm{grd}}_{\mathrm{dec}}$. Finally, the dual-domain U-Net based discriminator loss can be defined as follows:
\begin{align}
 \mathcal{L}_{D^{\mathrm{dud}}} = \mathcal{L}_{D^{\mathrm{img}}}+\mathcal{L}_{D^{\mathrm{grd}}}.
\end{align}

\subsection{CutMix Regularization}

The discriminator $D^{\mathrm{img}}$ suffers from the decreasing capability in recognizing the local differences between real and fake samples as the training goes, which may unexpectedly harm the denoising performance. Besides, the discriminator is supposed to focus on structure change at the global level and local details at the per-pixel level. To address these issues, we adopt the CutMix augmentation technique to regularize the discriminator inspired by~\cite{schonfeld2020u,yun2019cutmix}, which can empower the discriminator to learn the intrinsic difference between real and fake samples. Specifically, CutMix technique generates a new training image from two images by cutting patches from the one and pasting them to another. We define this augmentation technique in the context of LDCT denoising as follows:
\begin{align}
    \mat{I}_{\mathrm{mix}} =& \mathrm{mix}(\mat{I}_{\mathrm{ND}}, \mat{I}_{\mathrm{den}}, \mat{M})\notag\\
                           =& \mat{M} \odot \mat{I}_{\mathrm{ND}}+(1-\mat{M}) \odot \mat{I}_{\mathrm{den}},
\end{align}
where $\mat{M} \in\{0,1\}^{w \times h}$ is a binary mask controlling how to mix the NDCT and denoised images, and $\odot$ represents the element-wise multiplication. 

\begin{figure}[t]
    \centering
    \includegraphics[width=1\linewidth]{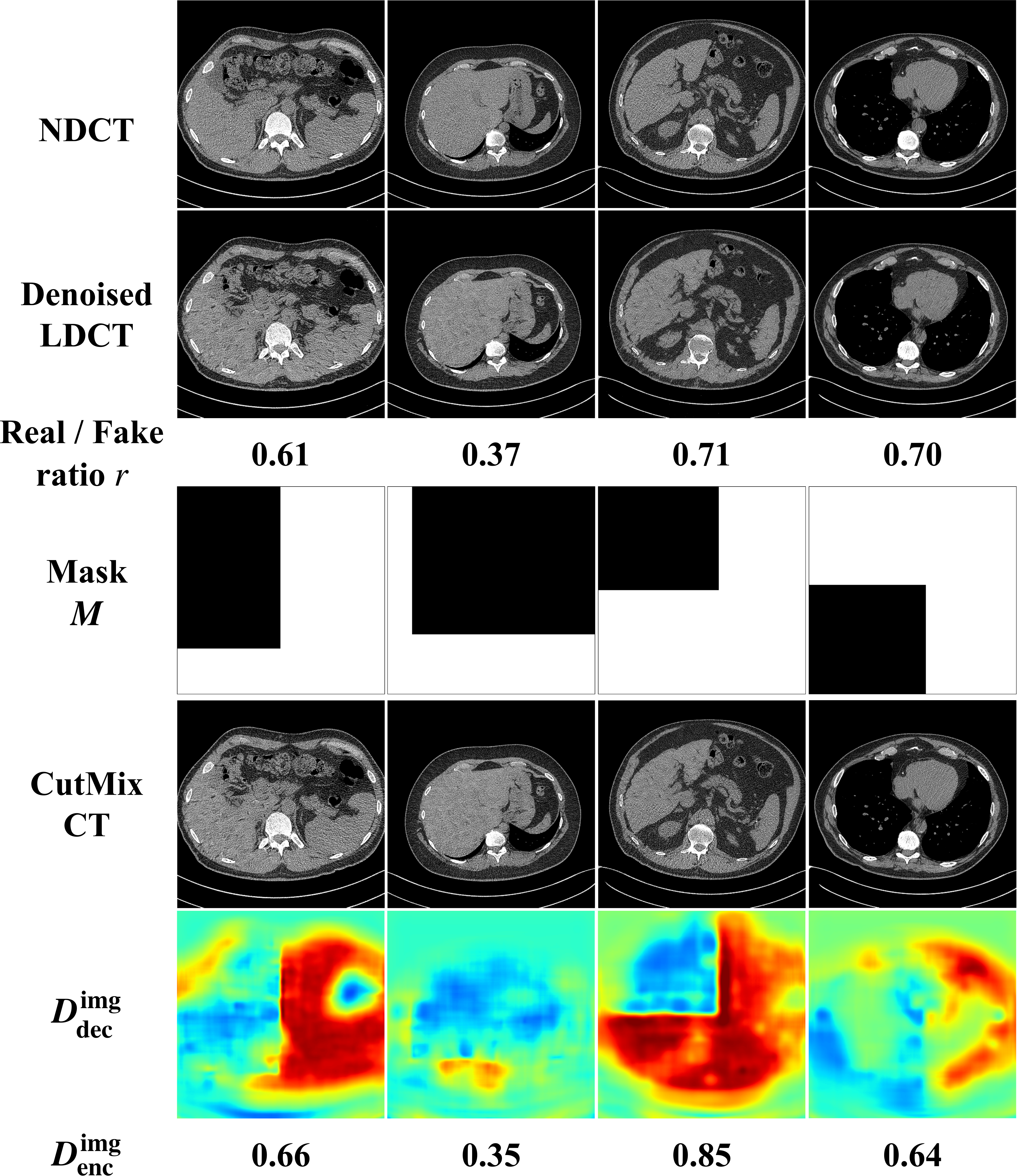}
    \caption{Illustration of the CutMix regularization for the U-Net based discriminator, \ie, $D^{\mathrm{img}}$. We randomly sample the ratio $r$ and top-left coordinates of the bounding box to form the mask $\mat{M}$ controlling where to crop. $D^{\mathrm{img}}_{\mathrm{dec}}$ is able to effectively capture the pixel differences between NDCT and denoised LDCT images while $D^{\mathrm{img}}_{\mathrm{enc}}$ can predict the mixed ratio. Note that the blue color  of $D^{\mathrm{img}}_{\mathrm{dec}}$ indicates the lower confidence score. Therefore, a well-trained discriminator can provide radiologists with a confidence map showing the uncertainty of the denoised results.}
    \label{fig:cutmix}
\end{figure}

The mixed samples should be regarded as fake samples globally by the encoder $D^{\mathrm{img}}_{\mathrm{enc}}$ since the CutMix operation has destroyed the global context of NDCT image; otherwise the CutMix may be introduced to denoised LDCT images during the training of GANs, causing undesirable denoising. Similarly, the $D^{\mathrm{img}}_{\mathrm{dec}}$ should be able to recognize the mixed area to provide the generator with accurate per-pixel feedback. Therefore, the regularization loss of CutMix can be formulated as:
\begin{align}
    \mathcal{L}_{\mathrm{reg}} = \mathbb{E}_{\mat{I}_{\mathrm{mix}}} \Big[[D^{\mathrm{img}}_{\mathrm{enc}}(\mat{I}_{\mathrm{mix}})]^{2} +  [D^{\mathrm{img}}_{\mathrm{dec}}(\mat{I}_{\mathrm{mix}})-\mat{M}]^{2}\Big],
\end{align}
where $\mat{M}$ used in CutMix also serves as the ground truth for $D^{\mathrm{img}}_{\mathrm{dec}}$.

Furthermore, to penalize the outputs of discriminator to be consistent with the per-pixel predictions after the CutMix operation, we further introduce another consistency loss following~\cite{schonfeld2020u} to regularize the discriminator with CutMix operation, which can be written as:
\begin{align}
    \mathcal{L}_{\mathrm{con}} = \| D^{\mathrm{img}}_{\mathrm{dec}} (\mat{I}_{\mathrm{mix}}) - \mathrm{mix}( D^{\mathrm{img}}_{\mathrm{dec}}(\mat{I}_{\mathrm{ND}}),  D^{\mathrm{img}}_{\mathrm{dec}}(\mat{I}_{\mathrm{den}}), \mat{M}) \|_F^2,
\end{align}
where $\|\cdot\|_F$ represents the Frobenius norm.

During training, the binary mask $\mat{M}$ is generated following the same pipeline as~\cite{yun2019cutmix,zhang2017mixup}. Specifically, we first sample the combination ratio $r$ from beta distribution $\mathrm{Beta}(1, 1)$ and then uniformly sample the top-left coordinates of the bounding box of cropping regions from $\mat{I}_{\mathrm{ND}}$ to $\mat{I}_{\mathrm{den}}$, with preserving the $r$ ratio. Similar to~\cite{yun2019cutmix,zhang2017mixup}, we employ a probability $p_{\mathrm{mix}}$ to control whether to apply the CutMix regularization technique for each mini-batch samples, which is empirically set to $0.5$. Fig.~\ref{fig:cutmix} presents the visual results of $D^{\mathrm{img}}$ with CutMix regularization technique. 
It can be observed that the outputs of $D^{\mathrm{img}}_{\mathrm{dec}}$ are the spatial combination of the real and generated patches with respect to the real/fake classification score. Therefore, the results have demonstrated the strong discriminative capability of the U-Net based discriminator in accurately learning per-pixel differences between real and generated samples, even though they are cut and mixed together to fool the discriminator. Besides learning the per-pixel local details, $D^{\mathrm{img}}_{\mathrm{enc}}$ can accurately predict the proportion of real patches, \ie, the mixed ratio, as it is to focus on the global structures.

\subsection{Network Architecture}

As we described above, our proposed method follows the GANs framework to optimize the generator effectively for LDCT denoising, with the U-Net based discriminator focusing on both global structures and local details, and an extra gradient branch encouraging better boundaries and details. In this subsection, we describe the network architectures of the generator and U-Net based discriminator.

\subsubsection{RED-CNN based generator}
In this paper, we employ RED-CNN~\cite{chen2017low} as the generator of our framework for LDCT denoising since this paper mainly focuses on the adversarial loss from dual-domain U-Net based discriminators. The main difference from~\cite{chen2017low} is that our framework is optimized in GANs manner, while the vanilla RED-CNN suffers the problem of over-smoothened LDCT images with MSE. Specifically, RED-CNN employs the U-Net architecture but removes the downsampling/upsampling operations to prevent information loss. We stack 10 (de)convolutional layers at both encoder and decoder, each of which has 32 filters for the sake of the computation cost, followed by a ReLU activation function. There are in total 10 residual skip connections.
It is important to note that although RED-CNN is adopted as the generator in our framework, the proposed method can be also adapted to other GANs-based methods such as CPCE~\cite{shan20183} and WGAN-VGG~\cite{yang2018low} with only changing the discriminators.

\subsubsection{U-Net based discriminator}
As detailed in Section~\ref{sec:unetD}, there are two independent discriminators in both image and gradient domains, each of which follows a U-Net architecture. Specifically, $D_{\mathrm{enc}}$ has 6 downsampling ResBlocks~\cite{he2016deep} with increasing number of filters; \ie 64, 128, 256, 512, 512, and 512. At the bottom of $D_{\mathrm{enc}}$, a fully-connected layer is used to output the global confidence score. Similarly, $D_{\mathrm{dec}}$ used the same number of ResBlocks in a reverse order to process the bilinearly upsampled features and the skip residuals of the same resolution, followed by a $1\times 1$ convolutional layer to output the per-pixel confidence map. Most importantly, a spectral normalization layer~\cite{miyato2018spectral} and a Leaky ReLU activation with a slope of 0.2 for negative input follow each convolutional layer of $D$ except the last one. 

We note that the network architectures of the generator and discriminator were proposed in literature; we did not propose a new network architecture to achieve the performance gain. One of our key contributions is to use the U-net as the discriminator in dual-domain to capture both local details and global structures for LDCT denoising.

\subsection{Loss Functions}\label{sec:losses}

\subsubsection{Adversarial loss}
Here we employ the sum of these two branches as the adversarial loss, which is defined in the context of least-squares GANs as follows:
\begin{align}
    \mathcal{L}_{\mathrm{adv}}= \mathbb{E} \Big[&\underbrace{[D^{\mathrm{img}}_{\mathrm{enc}}(\mat{I}_{\mathrm{den}}\big)-1]^{2} +  [D^{\mathrm{img}}_{\mathrm{dec}}(\mat{I}_{\mathrm{den}})-1]^{2}}_{\text{image\ domain}}+\\
    &
    \underbrace{[D^{\mathrm{grd}}_{\mathrm{enc}}(\nabla(\mat{I}_{\mathrm{den}}))-1]^{2} +  [D^{\mathrm{grd}}_{\mathrm{dec}}(\nabla(\mat{I}_{\mathrm{den}}))-1]^{2}}_{\text{gradient\ domain}}\Big],\notag
\end{align}
where $\nabla$ denotes the Sobel operator to obtain the image gradient.

\subsubsection{Pixel-wise loss}
To encourage the generator output the denoised LDCT images that match the NDCT images with both pixel level and gradient level, we adopt an pixel-wise loss between the NDCT images and denoised LDCT images, which includes a pixel loss and gradient loss for each branch as shown in Fig.~\ref{fig:framework}. The additional gradient loss can encourage to better preserve edge information at the pixel level. The two losses can be written as:
\begin{align}
    \mathcal{L}_{\mathrm{img}} &= \mathbb{E}_{(\mat{I}_{\mathrm{LD}},\mat{I}_{\mathrm{ND}})}\Big\|\mat{I}_{\mathrm{den}}-\mat{I}_{\mathrm{ND}}\Big\|^2_F, \\
    \mathcal{L}_{\mathrm{grd}}&=\mathbb{E}_{(\mat{I}_{\mathrm{LD}},\mat{I}_{\mathrm{ND}})}\Big|\nabla(\mat{I}_{\mathrm{den}})-\nabla(\mat{I}_{\mathrm{ND}})\Big|.
\end{align}
Note that we employ the mean squared error in pixel level rather than the feature level using pretrained model~\cite{shan20183,yang2018low} for the sake of computation cost, and the absolute mean error in gradient level as the gradients is much sparser than pixels.

\subsubsection{Final loss}
To encourage the generator to generate photo-realistic denoised LDCT images with better edge information and alleviate streak artifacts, the final loss function to optimize the generator $G$ is expressed as:
\begin{align}
    \mathcal{L}_{G} = \lambda_{\mathrm{adv}} \mathcal{L}_{\mathrm{adv}} + \lambda_{\mathrm{img}} \mathcal{L}_{\mathrm{img}} + \lambda_{\mathrm{grd}} \mathcal{L}_{\mathrm{grd}},
    \label{eq:g_loss}
\end{align}
where $\lambda_{\mathrm{adv}}$, $\lambda_{\mathrm{img}}$ and $\lambda_{\mathrm{grd}}$ are the weights for $\mathcal{L}_{\mathrm{adv}}$, $\mathcal{L}_{\mathrm{img}}$ and $\mathcal{L}_{\mathrm{grd}}$, respectively. Here, we empirically determine the hyper-parameters in a sequential way.  First, with only pixel-wise loss, our proposed DU-GAN reduces to RED-CNN since discriminators are not included during training. Although fast convergence, only optimizing the MSE loss leads to over-smoothing and blurred results, causing the loss of structural details. We set $\lambda_{\mathrm{img}}$ to be 1. Second, we tune the $\lambda_{\mathrm{adv}}$ to control the importance of adversarial loss to capture the texture details. We start from a small value for $\lambda_{\mathrm{adv}}$, and then gradually increase the importance of the adversarial loss, and visualize the denoising results. Finally, we tune $\lambda_{\mathrm{grd}}$ to capture edge information with a large value as the gradients are much sparser than pixels.

The discriminators $D^{\mathrm{img}}$ and $D^{\mathrm{grd}}$ are optimized by minimizing the following mixed loss:
\begin{align}
    \mathcal{L}_{D} = \mathcal{L}_{D^{\mathrm{dud}}} + \mathcal{L}_{\mathrm{reg}} + \mathcal{L}_{\mathrm{con}}.
    \label{eq:d_loss}
\end{align}
Note that we employ the same loss function in~\eqref{eq:d_loss} to optimize both $D^{\mathrm{img}}$ and $D^{\mathrm{grd}}$ but they are independent to each other and $D^{\mathrm{grd}}$ has an additional Sobel operator to compute the gradients.

\subsection{Complexity of \methodname}
Next, we discuss the complexity of the \methodname in terms of hyper-parameters and computational costs. First, compared to MSE-based methods that directly optimize mean-squared loss, DU-GAN is a GANs-based method that introduces an additional adversarial loss to the training process. Compared to vanilla GANs-based methods with a traditional classification discriminator, DU-GAN proposed to use the U-Net based discriminator to focus on both local details and global structures. Furthermore, DU-GAN also introduces another gradient branch along the original pixel branch to encourage clear boundaries. Therefore, there is only one extra hyper-parameter to control the importance of the gradient branch. Second, the main computational costs of DU-GAN come from the proposed U-Net based discriminator and gradient branch. However, such computational costs are affordable considering the better denoising quality and performance for our DU-GAN and only happen during the training stage. That is, the inference efficiency is still the same as the traditional ones.

\section{Experiments}\label{sec:exp}

This section presents the datasets, implementation details, qualitative and quantitative evaluations, uncertainty visualization, and ablation study.

\begin{figure*}[t]
    \centering
    \includegraphics[width=1\linewidth]{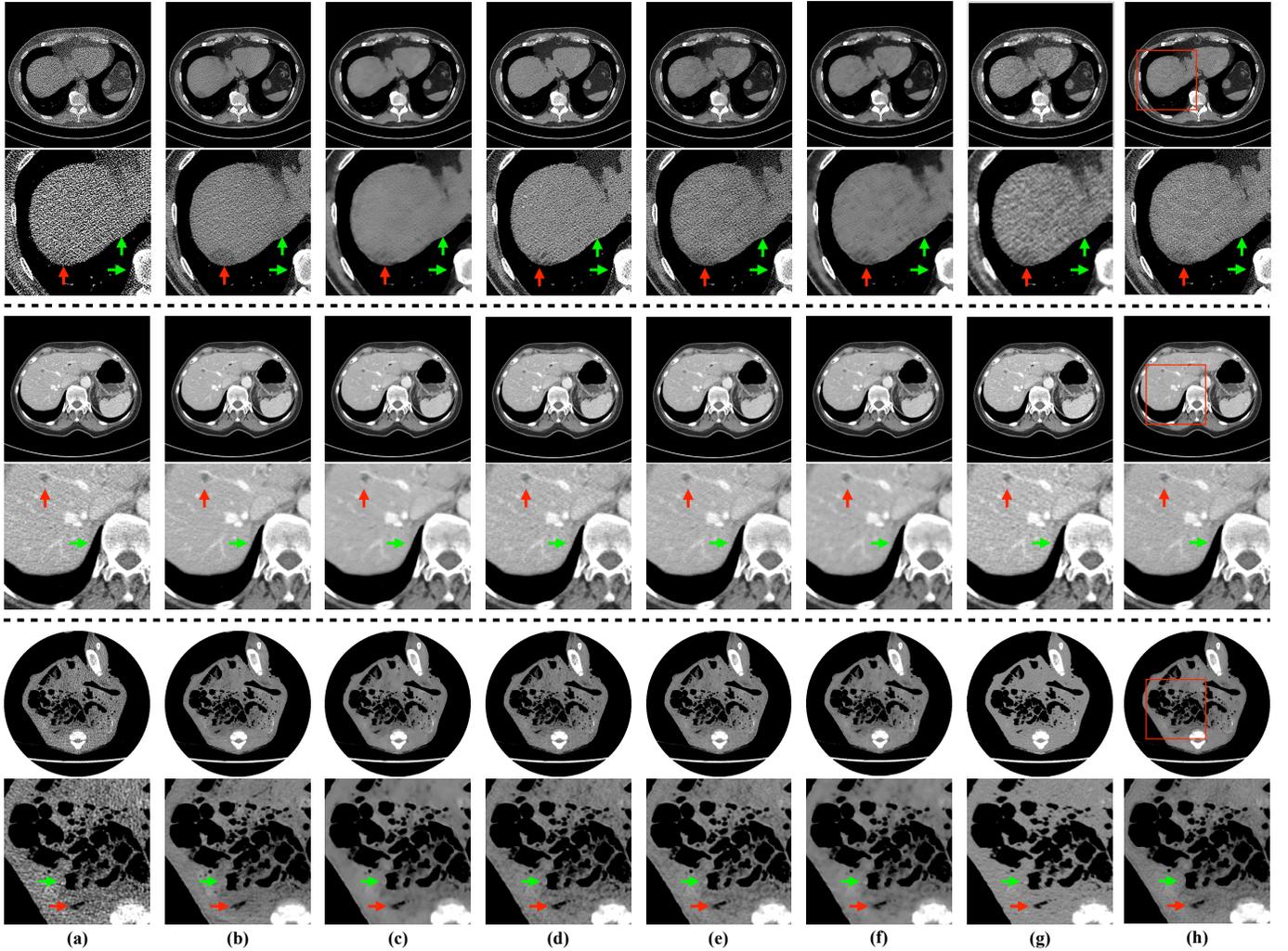}
    \caption{Transverse CT images from the Mayo-10\%, Mayo-25\%, and Piglet-5\%: (a) LDCT; (b) NDCT; (c) RED-CNN; (d) WGAN-VGG; (e) CPCE-2D; (f) Q-AE; (g) CNCL; and (h) \methodname (ours). Zoomed ROI of the red rectangle is shown below the full-size one. The display window is [-160, 240] HU for better visualization. Red arrow indicates low attenuation lesion. Green arrow indicates the white edge artifacts shown in some baseline algorithms while not shown in our method.}
    \label{fig:qualitative_comparisons}
\end{figure*}
\subsection{Datasets}

\subsubsection{Simulated dataset} The LDCT dataset used in this study was originally for~\emph{the 2016 NIHAAPM-Mayo Clinic Low-Dose CT Grand Challenge}, and lately released in~\cite{moen2021low}. It provides scans from three regions of the body with different \emph{simulated} low doses; \ie, head with $25\%$ of normal-dose, abdomen with $25\%$, and chest with $10\%$. In our experiments, we used the 25\% abdomen and 10\% chest datasets, named Mayo-25\% and Mayo-10\%, respectively. We evaluated our method on abdomen scans for comparisons with most previous works, and conducted experiments on chest scans since $10\%$ of normal-dose at chest is rather challenging compared to the $25\%$ of normal-dose at abdomen. For each dataset, we randomly select 20 patients for training and another 20 patients for testing; no identity overlapping between training and testing. In detail, 300K and 64K image patches were randomly selected from each set. For more information about this dataset, please refer to~\cite{moen2021low}.

\begin{figure*}[t]
    \centering
    \includegraphics[width=1\linewidth]{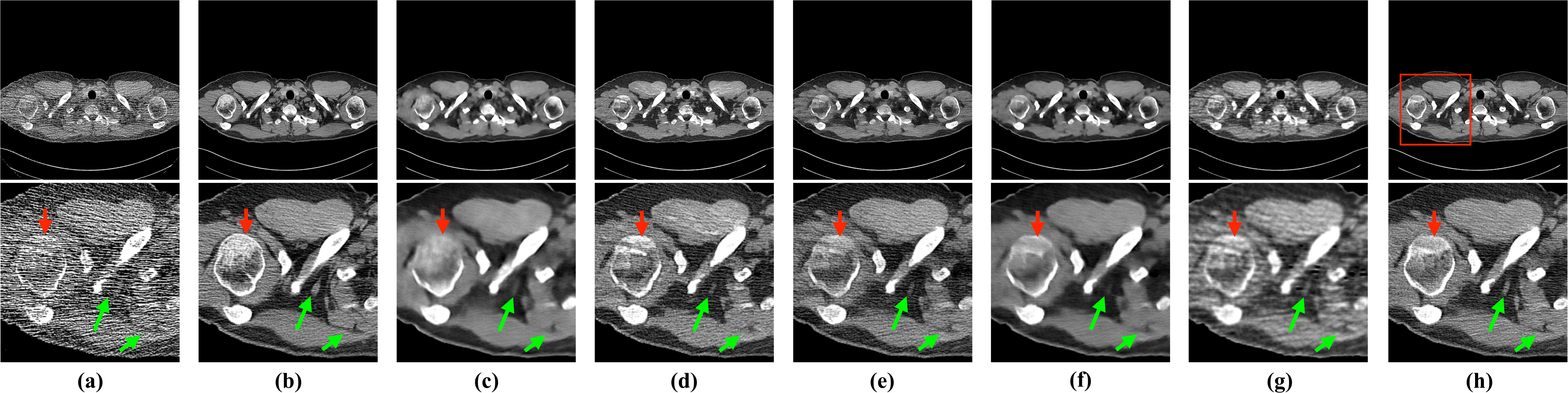}
    \caption{Transverse neck CT images from the Mayo-10\%: (a) LDCT; (b) NDCT; (c) RED-CNN; (d) WGAN-VGG; (e) CPCE-2D; (f) Q-AE; (g) CNCL; and (h) \methodname (ours). Zoomed ROI of the red rectangle is shown in the second row. The display window is [-160, 240] HU for better visualization. Red arrow indicates bone area while green arrow indicates a small structure.}
    \label{fig:qualitative_comparisons_artifacts}
\end{figure*}
\begin{table*}[t]
    \centering
    \caption{Quantitative comparisons of different methods on the testing sets of two simulated datasets and one real-world dataset. The best results among MSE- and GAN-based methods are marked in bold.}
    \label{tab:quantitative_comparisons}
    \begin{tabular*}{1\textwidth}{@{\extracolsep{\fill}}rlccccccccccc}
    \toprule
    & \multirow{2}{*}{Method} & \multicolumn{3}{c}{Mayo-10\%} & & \multicolumn{3}{c}{Mayo-25\%} & & \multicolumn{3}{c}{Piglet-5\%} \\ 
    \cmidrule{3-5} \cmidrule{7-9} \cmidrule{11-13} 
    & & PSNR$\uparrow$ & RMSE$\downarrow$ & SSIM$\uparrow$ & & PSNR$\uparrow$ & RMSE$\downarrow$ & SSIM$\uparrow$ & & PSNR$\uparrow$ & RMSE$\downarrow$ & SSIM$\uparrow$ \\ 
    \midrule
    &LDCT & 14.6382 & 0.1913 & 0.6561 & & 31.5517 & 0.0283 & 0.8639 & & 28.7279 & 0.0395 & 0.8587 \\ \hline
    \multirow{2}{*}{MSE-based}&RED-CNN~\cite{chen2017low} & \textbf{23.1388} & \textbf{0.0721} & \textbf{0.7249} & & 34.5740 & \textbf{0.0196} & \textbf{0.9236} & & 26.9691 & 0.0450 & \textbf{0.9318} \\
    &Q-AE~\cite{fan2019quadratic} & 21.3149 & 0.0884 & 0.7045 & & \textbf{34.6477} & 0.0197 & 0.9215 & & \textbf{29.7081} & \textbf{0.0331} & 0.9317 \\ \hline
    \multirow{3}{*}{GAN-based}&WGAN-VGG~\cite{yang2018low} & 20.3922 & 0.0992 & 0.7029 & & 33.2910 & 0.0226 & 0.9092 & & \textbf{30.3787} & \textbf{0.0318} & 0.9232 \\
    &CPCE-2D~\cite{shan20183} & 20.1435 & 0.0899 & 0.7295 & & 33.0612 & 0.0232 & 0.9125 & & 28.5329 & 0.0379 & 0.9211 \\ 
    &CNCL~\cite{geng2021content} & 21.8964 & 0.0852 & 0.7110 & & 32.4967 & 0.0243 & 0.9048 & & 28.5673 & 0.0383 & 0.9132 \\ 
    
    &DU-GAN (ours) & \textbf{22.3075} & \textbf{0.0802} & \textbf{0.7489}  && \textbf{34.6186} & \textbf{0.0196} & \textbf{0.9196} && 29.8598 & 0.0325 & \textbf{0.9345} \\ 
    \bottomrule
    \end{tabular*}
\end{table*}

\subsubsection{Real-world dataset} The real-world dataset from~\cite{yi2018sharpness} includes 850 CT scans of a deceased piglet obtained by a GE scanner (Discovery CT750 HD). The dataset provides CT scans of the normal-dose, 50\%, 25\%, 10\% and 5\% dose with a size of $512\times 512$, 708 of which is served for training while the left for testing. We evaluated our method on 5\% low-dose CTs as it is the most challenging dose, where the dataset is named Piglet-5\%. We randomly selected 60K and 12K image patches from training and testing sets, respectively. For more information about this dataset, please refer to~\cite{yi2018sharpness}.

\subsection{Implementation Details}
Following~\cite{shan20183,yang2018low,fan2019quadratic}, we employed the image patches with a size of $64\times 64$ and a window of $[-300, 300]$ to train all models with emphasis on tissue CT window, which are then directly applied to the whole image for visualization and testing. 
Note that we excluded those image patches that were mostly air. During training, all images are linearly normalized to $[0,1]$.

During training, we trained the model with a maximum of 100K iterations and with a mini-batch of size 64 on one NVIDIA V100 GPU. All networks in the proposed framework are initialized with He initialization~\cite{he2015delving}, and optimized by Adam optimization method~\cite{kingma2014adam} with a fixed learning rate of $10^{-4}$. The hyperparameters in the loss functions were empirically set as follows: $\lambda_{\mathrm{adv}}$ was $0.1$; $\lambda_{\mathrm{img}}$ was $1$; and $\lambda_{\mathrm{grd}}$ was $20$. We implemented four deep-learning-based methods including RED-CNN~\cite{chen2017low}, WGAN-VGG~\cite{yang2018low}, CPCE-2D~\cite{shan20183}, Q-AE~\cite{fan2019quadratic}, and CNCL~\cite{geng2021content} with the reference of official source code. 

\subsection{Qualitative Evaluations}

To demonstrate the effectiveness of the proposed method in generating photo-realistic denoised results with faithful details, 
Fig.~\ref{fig:qualitative_comparisons} showcases the representative results from three different datasets while Fig.~\ref{fig:qualitative_comparisons_artifacts} presents the results of one neck CT slice with strong streak artifacts.
The regions-of-interest~(ROIs) marked by the red rectangles are zoomed below, respectively.

\begin{figure*}[t]
    \centering
    \includegraphics[width=1.0\linewidth]{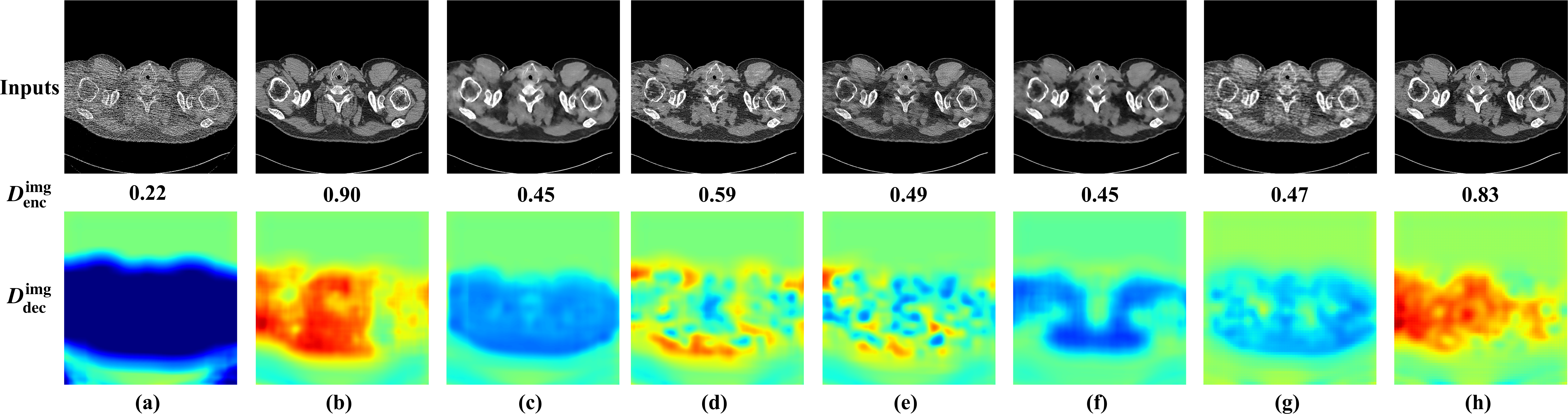}
    \caption{Uncertainty visualization of applying the trained discriminator to the outputs of different methods: (a) LDCT; (b) NDCT; (c) RED-CNN; (d) WGAN-VGG; (e) CPCE-2D; (f) Q-AE; (g) CNCL; and (h) \methodname (ours). The display window is [-160, 240] HU for better visualization. Note that the blue  color of $D^{\mathrm{img}}_{\mathrm{dec}}$ indicates the lower confidence score while red color indicates higher confidence score.}
    \label{fig:interpretability_visualization}
\end{figure*}

All methods present visually well denoised results to some degrees. However, RED-CNN and Q-AE over-smoothed and blurred the LDCT images as they are optimized by the MSE loss that tends to average the results, causing the loss of structural details. Although WGAN-VGG and CPCE-2D have greatly improved the visual fidelity, as expected, due to the use of adversarial loss, minor streak artifacts can still be observed since their traditional classification discriminator only provide the generator with global structure feedback.
Besides, they employed the perceptual loss in the high-level feature space to suppress the blurriness resulting from MSE loss. The perceptual loss, however, can only preserve the structures of NDCT images since some local details may be lost after processed by a pre-trained model. For example, the low attenuation lesions in Fig.~\ref{fig:qualitative_comparisons}, and the bones in Fig.~\ref{fig:qualitative_comparisons_artifacts} are less clear by WGAN-VGG and CPCE-2D while they can be easily observed in NDCT as well as the results of our methods. Most importantly, the small structures with their boundaries are consistently preserved with a clear visual fidelity.
This benefits from the well-designed dual-domain U-Net based discriminators, which can provide feedback of both global structures and local details to the generator, compared to the traditional classification discriminator used in WGAN-VGG and CPCE-2D with only structure information. Besides, the gradient domain branch can also encourage the denoising model to better preserve edge information.

Beyond encouraging better edge, Fig.~\ref{fig:qualitative_comparisons_artifacts} also demonstrates its impressive performance in dealing with the LDCT images with strong streak artifacts caused by photon starvation. Compared to the baseline methods that produce strongly blurry and ghosted denoised results, our method can effectively address this problem in the following aspects:
\begin{enumerate}
    \item streak artifacts can be easily detected by the gradient domain branch; and
     \item once detected, the dual-domain U-Net discriminators can fill the occlusion area by adversarial training to alleviate the impact of streak artifacts. 
\end{enumerate}
In summary, all of these  results further validate the superiority of our methods.

\subsection{Quantitative Evaluations}

For quantitative evaluations, we adopted three widely-used metrics including peak signal-to-noise ratio~(PSNR), structural similarity~(SSIM), and root mean square error~(RMSE). More specifically, PSNR and RMSE measure the denoising performance at pixel level while SSIM computes the structural similarity within a window. Table~\ref{tab:quantitative_comparisons} presents the results of different methods. First, RED-CNN and Q-AE are MSE-based denoising methods as they are directly trained with solely MSE loss. Although they achieve better PSNR and RMSE results, the visual results in Figs.~\ref{fig:qualitative_comparisons} and~\ref{fig:qualitative_comparisons_artifacts} confirm that MSE-based methods produce over-smoothed results compared to the NDCT images, leading to lose of structural information~\cite{shan20183,yang2018low,johnson2016perceptual}.  Note that the over-smoothed denoising results lead to a lower SSIM score. 
Second, WGAN-VGG, CPCE-2D, and our DU-GAN are GAN-based methods. 
CPCE-2D performed better than WGAN-VGG due to the conveying path since WGAN-VGG has to reconstruct the denoised results from the input LDCT images. Obviously, our method performs the best in terms of SSIM score with high visual fidelity while the PSNR and RMSE are also better than WGAN-VGG and CPCE-2D, indicating the superior denoising performance of our method while better structural fidelity.

Though \methodname used the same network architecture of the RED-CNN as the denoising model, their qualitative and quantitative  difference directly comes from the adversarial training and dual-domain U-Net based discriminators. The results of \methodname preserve more structural details that are important for diagnosis, at the cost of compromising the quantitative metrics such as PSNR and RMSE. We note that PSNR and RMSE are pixel-wise metrics, poorly correlating with human perception of image quality~\cite{goodfellow2014generative}.

\subsection{Uncertainty Visualization}

Fig.~\ref{fig:cutmix} shows the proposed discriminator with the U-Net architecture and CutMix regularization can robustly learn the per-pixel differences of local details between NDCT and denoised LDCT images by the decoder and also focus on the global structures by the encoder. With this well-trained discriminator, we can provide radiologist with a confidence map showing the uncertainty of the denoised results, since it is to learn the distribution of real samples, \ie, NDCT images. Therefore, we directly applied the trained discriminator $D^{\mathrm{img}}$ to the LDCT, NDCT, and the denoised LDCT images of different methods. 

Fig.~\ref{fig:interpretability_visualization} shows the uncertainty visualization. 
Obviously, the discriminator can accurately distinguish the LDCT from NDCT images, on both global score and per-pixel confidence. As both RED-CNN and Q-AE over-smoothen the LDCT images, the abdomen area of transverse CT image becomes better than LDCT images on the confidence map, according to the results of $D^{\mathrm{img}}_{\mathrm{dec}}$. This also explains why RED-CNN and Q-AE have the lowest global score of $D^{\mathrm{img}}_{\mathrm{enc}}$, which indicates that the discriminator can robustly detect the blurriness in the CT images.
Furthermore, although CPCE-2D can produce more clear denoised results than RED-CNN, the streak artifacts significantly compromised the quality of the denoised results.  Similarly, the WGAN-VGG has learned more local details than CPCE-2D but it still cannot handle the impact of  the streak artifacts. On the contrary, the proposed method can produce the most photo-realistic denoised results with the highest global score. Compared to the traditional classification discriminator used in CPCE-2D and WGAN-VGG, our \methodname can provide the generator with the per-pixel feedback by learning the local detail differences. It can be seen from the per-pixel of $D^{\mathrm{img}}_{\mathrm{dec}}$. In other words, we achieve a more smooth per-pixel confidence, indicating that the discriminator cannot distinguish the real and fake samples at the per-pixel level.

\subsection{Ablation Study}
\label{sec:ablation_study}

In this subsection, we conducted the ablation study of our method to fully explore the proposed method in terms of the importance of different components, the architectures of discriminator, and the different patch sizes. The ablation study was done on the testing set of Mayo-10\% dataset, which includes a total of 6,590 slices from 20 patients.

\subsubsection{Components analysis}
We investigate the impact of the U-Net based discriminator in the image domain, CutMix regularization, and dual-domain training (\ie, with gradient branch) by gradually applying them to the baseline method. Similar to WGAN-VGG and CPCE-2D, the baseline method only includes the traditional classification discriminator with the same hyperparameters for fair comparison.

\begin{table}[h]
\centering
    \caption{Ablation study of component analysis. Our method is the baseline method with U-Net discriminator in the image domain, CutMix regularization, and the U-Net discriminator in the gradient domain (dual-domain).  The best results are marked in bold.}\label{tab:component_analysis}
        \begin{tabular*}{1\linewidth}{@{\extracolsep{\fill}}lccc}
            \toprule
            Method  & PSNR$\uparrow$  & RMSE$\downarrow$ & SSIM$\uparrow$ \\
            \midrule
            Baseline   & 21.4988 & 0.0871 & 0.7365 \\
            \quad + U-Net Based Discriminator  & 22.1214 & 0.0816 & 0.7454 \\
            \quad \quad + CutMix Regularization   & 21.7894 & 0.0844 & 0.7477 \\
            Ours~(+ Dual-Domain)   & \textbf{22.3075} & \textbf{0.0802} & \textbf{0.7489} \\
            \bottomrule
            \end{tabular*}
\end{table}

Table~\ref{tab:component_analysis}
presents the quantitative results for ablation study. 
First, replacing the traditional classification discriminator with a U-Net based discriminator can simultaneously provide the generator with both global structure and local per-pixel feedback, which leads to a significant increase in terms of SSIM. Second, when we further use CutMix technique to regularize the U-Net based discriminator, the mixed samples can boost the discriminant capacity of discriminator and make discriminator more focus on the local details, leading to the increased SSIM score and a slightly decreased PSNR and RMSE.
Last, further adding the U-net based discriminator in the gradient domain  into the method above forming the dual-domain training yields our method. Specifically, the additional gradient domain training can help our method remove the streak artifacts and encourage more clear edge in the denoised LDCT images. As a result, it can effectively improve all metrics including the PSNR and RMSE in pixel space and SSIM in visual similarity.

\begin{table}[h]
\centering
    \caption{Ablation study of different discriminators on testing set of Mayo-10\% dataset. The best results are marked in bold.}\label{tab:discriminator_architecture}
        \begin{tabular*}{1\linewidth}{@{\extracolsep{\fill}}cccc}
            \toprule
            Method  & PSNR$\uparrow$  & RMSE$\downarrow$ & SSIM$\uparrow$ \\
            \midrule
            Patch   & 21.4988 & 0.0871 & {0.7365} \\
            Global  & {22.6810} & {0.0760} & 0.7262 \\
            Pixel   & \textbf{23.1102} & \textbf{0.0724} & 0.7343 \\
            U-Net   & 22.1214 & 0.0816 & \textbf{0.7454} \\
            \bottomrule
            \end{tabular*}
            
\end{table}
\subsubsection{Architectures of discriminator} Since the architectures of discriminator play a critical role in the training of GANs, it is worthwhile studying the advantage of the U-Net based discriminator over other classical discriminator architectures such as patch discriminator~\cite{isola2017image}, pixel discriminator~\cite{isola2017image}, and traditional global discriminator. Compared to traditional classification discriminator that classifies the real and fake samples at image level, patch discriminator focuses on the image patches. Due to the patch training of low-dose CT denoising, this discriminator architecture can be seen as the patch discriminator. The discriminator with seven convolutional layers and one fully-connected layer is regarded as the global discriminator. On the other hand, the pixel discriminator~\cite{isola2017image} contains 7 $1\times 1$ convolutional layers to penalizes the generator at per-pixel level. For fair comparisons, we trained patch and pixel discriminator with image patches and trained the global discriminator with the whole images with the size of $512\times 512$, respectively. Table~\ref{tab:discriminator_architecture} shows that the combination of global and pixel information in U-Net based discriminator produces the best SSIM score. This indicates the advantage of U-net-based discriminator for LDCT denoising over other classical discriminator architectures such as patch discriminator, pixel discriminator, and traditional global discriminator. Instead of pixel discriminator only capturing per-pixel difference and traditional classification discriminator only focusing on global structure, the U-Net based discriminator has the advantages of both worlds, yielding better quantitative results and denoising quality.

\begin{table}[t]
\centering
\caption{Ablation study of patch sizes on the testing set of Mayo-10\% dataset. The best results are marked in bold.}\label{tab:patch_size}
\begin{tabular*}{1\linewidth}{@{\extracolsep{\fill}}cccc}
    \toprule
    Size        & PSNR$\uparrow$  & RMSE$\downarrow$ & SSIM$\uparrow$ \\
    \midrule
    64$\times$64   & \textbf{22.3075} & \textbf{0.0802} & \textbf{0.7489} \\
    128$\times$128  & {22.0218} & {0.0826} & {0.7479} \\
    256$\times$256   & 21.8478 & 0.0843 & 0.7467 \\
    512$\times$512   & 21.7254 & 0.0855 & 0.7441 \\
    \bottomrule
\end{tabular*}
\end{table}
\subsubsection{Patch size}
Due to the U-Net architecture of the discriminator, it is also important to analyze the influence of the patch size during training. However, it is very difficult to directly train the denoising model from scratch. Therefore, we trained our model with the image size of $64\times 64$, $128\times 128$, $256\times 256$, and $512\times 512$, and we fine-tuned the generator based on the model trained on previous smaller size. Table~\ref{tab:patch_size} shows that a small patch size can achieve better performance because the larger patch sizes may introduce training difficulties with less training samples.

\section{Discussion and Conclusion}\label{sec:conc}

In this paper, we proposed a novel \methodname for LDCT denoising. The introduced U-Net based discriminator can not only provide the per-pixel feedback to the denoising network but also focus on the global structure. We further add an extra U-Net based discriminator into the gradient domain, which can enhance the edge information and alleviate the streak artifacts caused by photon starvation. We also examined that the CutMix technique can boost the training of discriminator, which can provide the radiologists with a confidence map on the uncertainty of the denoised results. Extensive experiments demonstrated the effectiveness of the proposed method through visual comparison and quantitative comparison.

Although a different architecture or  model size could certainly affect results, our DU-GAN has demonstrated its generalization ability on two simulated low-dose CT datasets of different doses and one real-world dataset, with the same network architecture and hyperparameters.  Our ablation study validates each component and their relative importance of all components should be consistent on a new dataset, which indicates that our DU-GAN can be easily adapted to different scenarios.

We acknowledge some limitations in this work. First, we used the qualitative and quantitative comparisons to evaluate the image quality. A human reader study may be needed to further validate its potential in clinical diagnosis, though there are significant difference between the proposed and other baseline methods. Second, the U-net based discriminator can provide radiologists with a confidence map of the denoised images. How this helps radiologists in clinical routine could be examined with specific tasks such as liver lesion diagnosis, which can be further studied as a future direction. Third, DU-GAN could introduce slightly more computational cost during training since it employs the U-Net as the discriminator and adopts a dual-domain training strategy. However, we emphasize that the extra computational cost is relatively affordable as DU-GAN trains the whole framework based on the $64\times 64$ image patches instead of the original image size $512\times 512$. Therefore, we believe that the computational cost can be significantly reduced. We note that the extra computational cost only happens for the training stage. That is, the inference efficiency is still the same as the traditional one as the the dual-domain discriminators are not involved during testing stage. Finally, we only validated \methodname with two dose levels in this paper and it is worth further validating \methodname with lower radiation dose.

In conclusion, the proposed \methodname achieves better denoising performance than other GAN-based models and has great potential for clinical use with uncertainty visualization.


\end{document}